\documentclass[prl,twocolumn,showpacs,superscriptaddress]{revtex4-1}
\usepackage{graphicx}
\usepackage{units}
\usepackage{multirow}
\usepackage{bigstrut}
\usepackage{overpic}
\usepackage{array}
\usepackage{color}
\usepackage{amsmath}
\usepackage{bm}
\usepackage{times}
\usepackage{adjustbox}
\RequirePackage{lineno}
\usepackage[absolute,overlay]{textpos}

\usepackage{changes}
\usepackage{appendix}
\usepackage[normalem]{ulem}
\usepackage[bookmarksnumbered, pdfstartview=FitH,colorlinks,urlcolor=blue, citecolor=blue,linkcolor=blue] {hyperref}%这里是让文献显示有颜色
\newcommand{\br}[1]{\mathcal{B}(#1)}

\newcommand{\wv}[1]{$\mathcal{#1}$-wave}

\newcommand{\fracDwave}{0.16   \pm  0.05_{\rm stat} \pm  0.02_{\rm syst}}
\newcommand{\rV}       {1.41  \pm 0.05_{\rm stat} \pm 0.01_{\rm syst}}
\newcommand{\rtwo}     {0.77  \pm 0.04_{\rm stat} \pm 0.02_{\rm syst}}

\newcommand{\BFnormal} {7.878\pm0.063_{\rm stat.} \pm 0.048_{\rm syst.}}

\newcommand{\BFKstarD}  {7.603\pm2.457_{\rm stat.} \pm 0.194_{\rm syst.}}
\newcommand{\BFSwave}  {0.462\pm0.015_{\rm stat.} \pm 0.016_{\rm syst.}}
\newcommand{\BFPwave}  {7.403\pm0.061_{\rm stat.} \pm 0.048_{\rm syst.}}
\newcommand{\BFDwave}  {1.265\pm0.409_{\rm stat.} \pm 0.032_{\rm syst.}}

\newcommand{\Nsig}{$28900\pm224$}
\newcommand{\EAsig}{$(24.90\pm0.02)\%$}

\newcommand{\BESIIIorcid}[1]{\href{https://orcid.org/#1}{\hspace*{0.1em}\raisebox{-0.45ex}{\includegraphics[width=1em]{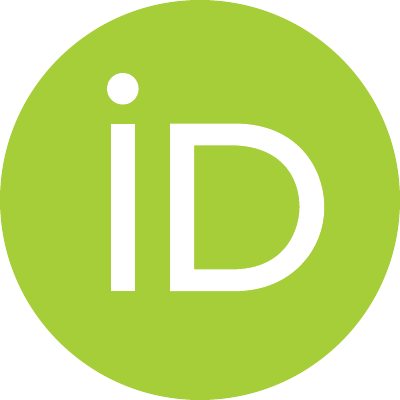}}}}

\begin{document}

\modulolinenumbers[1]
%\linenumbers

%\setlength{\oddsidemargin}{-0.5cm} \addtolength{\topmargin}{15mm}

\title{\boldmath First Amplitude Analysis of $D^0\rightarrow K^-\pi^0e^+\nu_e$ and Observation of $D^0\rightarrow K^*_2(1430)^-e^+\nu_e$  }

\author{
M. Ablikim \emph{et al.$^{*}$}\\
(BESIII Collaboration)\\
}
\date{February 28, 2026}
%\include{authorlist}
%%%%%%%%%%%%%%%%%%%%%%%%%%%%%%%%%%%%%%%%%%%%%%%%%%%%%%%%%%%%%%%%%%%%%%%%%%%%%%%%%%%%%%%%%%

\begin{abstract}
We present the first amplitude analysis of the semileptonic decay $D^0\to K^-\pi^0 e^{+}\nu_{e}$  by analyzing  $e^+e^-$ annihilation data corresponding to an integrated luminosity of 20.3 fb$^{-1}$ collected at the center-of-mass energy of 3.773 GeV with the BESIII detector. A tiny \wv{D} component of the $K^*_2(1430)^-$  accounting for $(\fracDwave)\%$ of the $K^-\pi^0$ is observed for the first time with a significance of $7.9\sigma$ in addition to the dominant \wv{P} component of $K^*(892)^-$ and the sub-dominant $K^-\pi^0$ \wv{S}. The hadronic form factors of the $D^0 \to K^*(892)^-$ transition are measured precisely as $r_V=V(0)/A_1(0)=\rV$ and $r_2=A_2(0)/A_1(0)=\rtwo$. The branching fraction of  $D^0\to K^*(892)^-e^+\nu_e$ with $K^*(892)^-\to K^-\pi^0$ is measured to be $(\BFPwave)\times10^{-3}$. Combining the measurements of the $D^0\to K^*(892)^-(K^*(892)^-\to K^-\pi^0)\ell^+ \nu_\ell$, lepton flavor universality is tested by the ratio $\mathcal{R}_{\rm LFU}=\mathcal{B}(D^0\to K^*(892)^-\mu^+ \nu_\mu)/\mathcal{B}(D^0\to K^*(892)^-e^+\nu_e)=0.928\pm0.020_{\rm stat}\pm0.012_{\rm syst}$ with unprecedented precision; no violation is found. Furthermore, isospin symmetry in the decay $K^*(892) \to K\pi$ is tested by $\mathcal R_{K^{*-}} =\mathcal{B}(K^*(892)^-\to K^- \pi^0)/\mathcal{B}(K^*(892)^-\to K_S^0 \pi^-)= 1.09\pm0.02_{\rm stat}\pm0.02_{\rm syst}$ for the first time using the previous measurement of $D^0\to K^*(892)^-e^+\nu_e$ with $K^*(892)^-\to K^0_S\pi^-$.
Finally, the phase shift of the $K\pi$ \wv{S} is extracted in a model-independent way, which sheds light on the nature of the lightest strange scalar meson, the $K^*_0(700)$.
\end{abstract}

\maketitle

% -------------------------- 核心：无geometry，自动左下角+短横线+正常边距 --------------------------
% 用revtex4-1原生的\oddsidemargin（左页边距）计算水平位置，无需硬编码
\begin{textblock*}{6cm}(\oddsidemargin+2.5cm, \paperheight-2.0cm)
  \footnotesize % PRL脚注字号
  \raggedright  % 左对齐
  % 短横线：长度3cm（不太长），线宽0.4pt（PRL标准线宽）
  \hrule width 3cm height 0.4pt \vspace{0.5em} 
  \textbf{\ \ \ \ \ *Full author list given at the end of the Letter.} \\
\end{textblock*}

%\begin{CJK}{UTF8}{gbsn}
%--------引言-----------引言------------------引言-----------------
Semileptonic (SL) decays of charmed mesons provide a good platform to study the non-perturbative regime of quantum chromodynamics (QCD), test lepton flavor universality (LFU), and extract Cabibbo-Kobayashi-Maskawa (CKM) matrix elements~\cite{ANTONELLI2010197,CKMDToV2025}. 
In particular, the four-body SL decay $D \to K\pi \ell \nu_\ell$  (denoted as $D_{\ell 4}$ ) has been the subject of much theoretical work~\cite{CQM2000,HQEFT2003,ABADA2003625,HMChiT2005,LCSR2006,CLFQM2012,CLFQM2012,LEChiQM2014,CLFQM2017,CQM2017,CCQM2019,LCSR2020,RQM2020,RQM2020,HQCD2024,LCSR2025} and experimental measurements\cite{MARK31991D02kpienu,FOCUS2002Dp2kpmunu,FOCUS2005D02kpimunu,PhysRevD.83.072001,BESIIIanffDp2kpienu,BESIIItaoluyanDp2kspienu, BESIIILiLeiDtoKpienu,BESIIIDtoKmpi0munu,BESIIID02kspimunu,BESIIIXiechengDp2kspilnu}. These decays offer an excellent opportunity to directly test theoretical methods with precise measurements, thereby deepening our understanding of the Standard Model (SM). 
They also contain abundant dynamical information on the $K\pi$ system, making them well-suited for studying strange mesons such as the $K^*_0(700)$ (also known as $\kappa$ ), $K^*(892)$, and $K^*_2(1430)$~\cite{PhysRevD.110.030001}, and so on. However, the decay $D^0 \to K^- \pi^0 e^+ \nu_e$  was only searched by MARK-III without observation in 1991\cite{MARK31991D02kpienu}. Now, it is time to study this decay with a large dataset at BESIII~\cite{BESIII3770data1,BESIII3770data2}.

%--------Dwave-----------引言-----------------Dwave-----------------
The tensor mesons, $K^*_2(1430)$ and $a_2(1320)$, as well as  the two isosinglet mesons, $f_2(1270)$ and $f^\prime_2(1525)$, collectively form the $1^3P_2$ nonet~\cite{PhysRevD.110.030001} in the quark model. Recently,  using SU(3) flavor symmetry, the branching fractions (BFs) of  $D^0 \to K^*_2(1430)^- \ell^+ \nu_\ell$ have been predicted to be $(1.36 \pm 0.49) \times 10^{-5}$ and $(0.91 \pm 0.31) \times 10^{-5}$ for positron and muon channels, respectively~\cite{BFKstar2WRM}. 
In comparison, the corresponding predictions from the relativistic quark model (RQM) are $(1.26 \pm 0.14) \times 10^{-5}$ and $(0.81 \pm 0.09) \times 10^{-5}$~\cite{TensorRQM}.
Although numerous analyses have been conducted on the $D_{\ell 4}$ decays,  there was still no observation for $K^*_2(1430)$ in the previous studies~\cite{BESIIIDtoKmpi0munu,PhysRevD.83.072001,BESIIIanffDp2kpienu,BESIIID02kspimunu, BESIIILiLeiDtoKpienu,BESIIItaoluyanDp2kspienu,BESIIID02kspimunu,BESIIIXiechengDp2kspilnu}. 
Any observation of $K^*_2(1430)$ in the $D_{\ell 4}$ decays is crucial to test these predictions. 

%--------SPwave-----------引言-----------------SPwave-----------------
The \wv{P} component of $K^*(892)$ dominates the $K\pi$ system in the $D_{\ell 4}$ decays~\cite{ FOCUS2002Dp2kpmunu,FOCUS2005D02kpimunu,BESIIIDtoKmpi0munu,PhysRevD.83.072001,BESIIIanffDp2kpienu,BESIIID02kspimunu, BESIIILiLeiDtoKpienu,BESIIItaoluyanDp2kspienu,BESIIID02kspimunu,BESIIIXiechengDp2kspilnu}. 
Many theoretical calculations on $D^0\to K^*(892)^-\ell^+\nu_\ell$ have been performed by various QCD models including Lattice QCD (LQCD)~\cite{ABADA2003625}, light-cone sum rules (LCSR)~\cite{LCSR2006,LCSR2020,LCSR2025},  constituent quark model (CQM)~\cite{CQM2017}, heavy quark effective theory (HQET)~\cite{HQEFT2003},  combined heavy meson and chiral theory (HM$\chi$T)~\cite{HMChiT2005}, covariant light-front quark model (CLFQM)~\cite{CLFQM2012,CLFQM2017},  large energy chiral quark model (LE$\chi$QM)~\cite{LEChiQM2014}, covariant confined quark model (CCQM)~\cite{CCQM2019} and RQM~\cite{RQM2020}. 
However, there are still significant discrepancies on the form factor (FF) predictions among these models. 
Therefore, precision measurements of these FFs are desirable. 
%Moreover, the  $D_{\ell 4}$ decays provide an independent determination of  the $c\to s$ CKM matrix element $|V_{cs}|$ to improve its precision. \cite{CKMDToV2025}. 

Previously, there were 2-3 $\sigma$ tensions~\cite{LFUBabar2012,LFUBabar2013,LFULHcb2014,LFULHcb2015,LFULHcb2017} reported in various LFU tests in SL $B$ decays by the LHCb, Belle and BABAR experiments.
%This LFU violation may be explained by the two-Higgs-doublet model~\cite{LFUTH2007,LFUTH2015} or the seesaw mechanism~\cite{LFUTh2012}. 
Precision measurements of the ratio $\mathcal R_{\rm LFU}=\frac{\br{D^0 \to K^*(892)^-\mu^+\nu_\mu}}{\br{D^0 \to K^*(892)^-e^+\nu_e}}$ in the $D_{\ell 4}$ decays provide critical and complementary tests of LFU in the weak interaction and could reveal new physics effects.
Additionally, it may be surprising that there is no measurement of the BFs of $K^*(892)^- \to K^- \pi^0 $ and $K^*(892)^- \to K_S^0 \pi^- $ since the $K^*(892)$ discovery in 1961~\cite{kpi1961}.  The value is usually taken as $\simeq 1/3 $, ignoring isospin-breaking effects. Theoretically, these effects are naively expected to be at most a few percent, but this has never been tested experimentally. 
The SL decay $D^0\to K^*(892)^-\ell^+ \nu_\ell$ provides a unique arena to perform the test via the ratio $\mathcal R_{K^{*-}}=\frac{\mathcal{B}(K^*(892)^- \to K^-\pi^0)}{\mathcal{B}(K^*(892)^- \to K_S^0\pi^-)}$ $=\frac{\mathcal{B}(D^0\to K^*(892)^- \ell^+\nu_\ell) \, \mathcal{B}(K^*(892)^-\to K^- \pi^0)}{\mathcal{B}(D^0\to K^*(892)^- \ell^+\nu_\ell) \, \mathcal{B}(K^*(892)^-\to K_S^0 \pi^-)}$. 

Finally, the $K^-\pi^0$ system contains a significant  \wv{S} component, whose properties can shed light on the nature of scalar mesons, particularly the long-standing puzzle of the $K^*_0(700)~$\cite{Yao-2020bxx}.
This resonance is a member of the lightest scalar meson nonet, together with the $f_0(500)$ (also known as $\sigma$), $f_0(980)$, and $a_0(980)$~\cite{Jaffe-2004ph, Pennington-2010dc, kappaFDR2016, Yao-2020bxx}. 
The pole parameters of the $K^*_0(700)$ have been extracted through dispersion relations~\cite{Zhou-2006wm,Descotes-Genon-2006sdr,kappaLQCD2012}, based on the only two datasets of the phase shift~\cite{LASS1988,Estabrooks-1977xe}. 
Thus, an additional measurement of the $K\pi$ phase shift in a clean environment is also valuable to constrain the pole parameters of the $K^*_0(700)$.

%--------This letter------This letter----------This letter--------------
This Letter reports the first dynamics study and BF measurement of  $D^0 \to K^- \pi^0 e^+ \nu_e$  using 20.3 fb$^{-1}$ of $e^+e^-$ collision data collected at the center-of-mass energy of $\sqrt{s}=3.773~\rm GeV$ with the BESIII detector. An amplitude analysis is used to decompose the $K^-\pi^0$ system into the $K^*_0(700)^-$ and  $K^*_0(1430)^-$ \wv{S} components, the  $K^*(892)^- $ \wv{P} component, the $K^*(1410)^- $ \wv{P^\prime} component, and the $K^*_2(1430)^-$ \wv{D} component. Charge-conjugate modes are implied throughout this Letter.

%---------------MC描述-------------MC描述-------------MC描述-------------
A detailed description of the design and performance of the BESIII detector is available in Refs.~\cite{Ablikim:2009aa,detector}. An inclusive Monte Carlo (MC) sample, as outlined in Refs.~\cite{BESIIIanffDp2kpienu}, is utilized to determine selection efficiencies and estimate background contributions. This inclusive MC sample includes the production of $D\bar{D}$, non-$D\bar{D}$ decays of the $\psi(3770)$, the initial state radiation production of the $J/\psi$ and $\psi(3686)$, and continuum processes.
In the generation of the signal, $D^0\rightarrow K^{-}\pi^0e^+\nu_e$, we take into account the knowledge of the amplitude results obtained in this work. 

%-------DT metho----------DT method ----------DT method-----------------
Near open-charm threshold at $\sqrt{s}=3.773~\rm GeV$, $D\bar{D}$ mesons are produced in pairs. The $\bar{D}^0$ (single-tag, ST) is fully reconstructed first in three channels $\bar{D}^0 \to K^+ \pi^-$, $\bar{D}^0 \to K^+ \pi^- \pi^0 $, and $\bar{D}^0 \to K^+ \pi^-\pi^- \pi^+$. In the presence of the ST $\bar{D}^0$, the $D^0\to K^- \pi^0e^+ \nu_e$ decay is reconstructed from the remaining tracks recoiling against the $\bar{D}^0$ meson to form double-tag (DT) candidates. More details of this DT method are given in Appendix A. Detailed descriptions of the selection criteria for  $\pi^\pm$, $K^\pm$, $e^\pm$, $\gamma$, and $\pi^0$ candidates are provided in Ref.~\cite{BESIIIanffDp2kpienu}.

%-----------ST selection-------------ST selection-----------ST selection-------
To separate ST $\bar{D}^0$ mesons from combinatorial backgrounds, a beam-constrained mass defined as $M_{\rm BC}= \sqrt{(\sqrt{s}/2)^2/c^4- |\vec{p}_{\bar{D}^0}|^2/c^2}$ is used, where $\vec{p}_{\bar{D}^0}$ represents the momentum of the $\bar{D}^0$ meson candidate. To improve signal significance, the kinematic variable $|\Delta E| = |E_{\bar{D}^0} - \sqrt{s}/2|$ is required to be within a certain range for each mode. Here $E_{\bar{D}^0}$ is the energy of the $\bar{D}^0$ candidate. For each ST mode and charm, if multiple  $\bar{D}^0$ candidates are found in an event, only the one with the minimum $|\Delta E|$ is retained for further analysis.
The $\bar{D}^0$ meson yields in each mode are determined through a binned maximum likelihood fit to the $M_{\rm BC}$ distributions. The signal is modeled using the MC simulated shape convolved with a double Gaussian function, while an ARGUS function describes the combinatorial  background. The total $\bar{D}^0$ yield is $N_{\rm ST}^{\rm tot} = (15375 \pm4)\times 10^3$. Details of $|\Delta E|$ and $M_{\rm BC}$ requirements,  efficiencies, and yields in each ST mode are listed in Supplemental Material~\cite{This-supply}.

% -----------------------DT selection ---------————DT selection-------------DT selection

\begin{figure}[htp]
    \centering
    \includegraphics[width=0.23\textwidth]{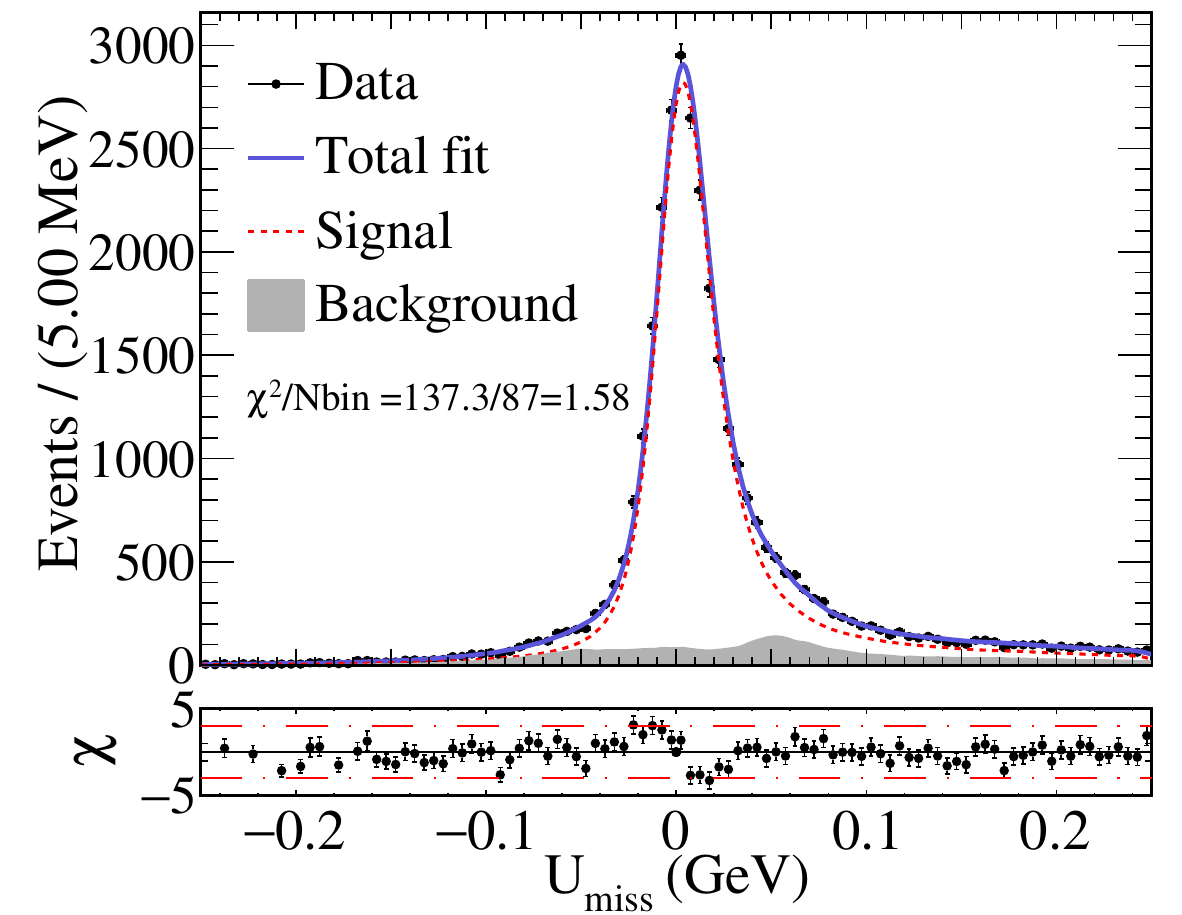} 
    \includegraphics[width=0.23\textwidth]{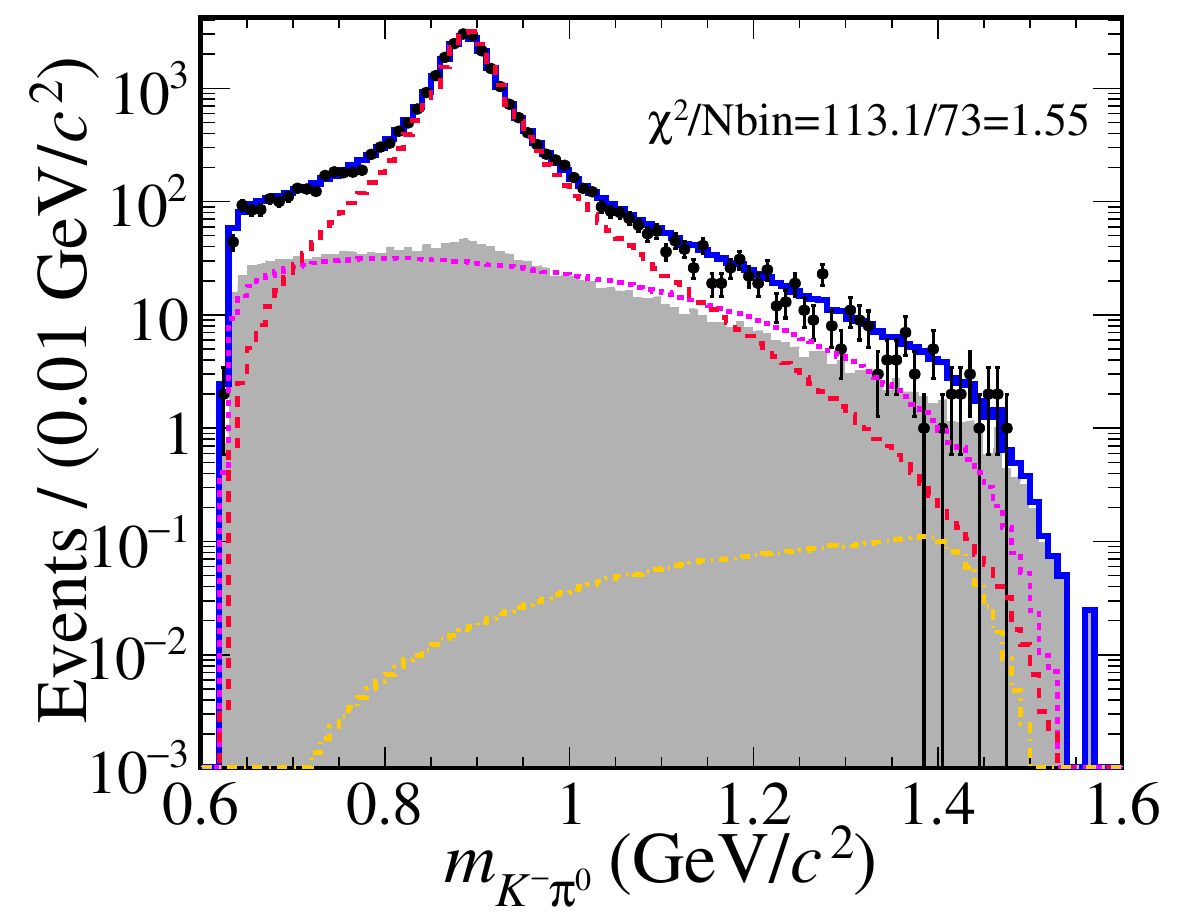}\\[0.1cm]
    \includegraphics[width=0.23\textwidth]{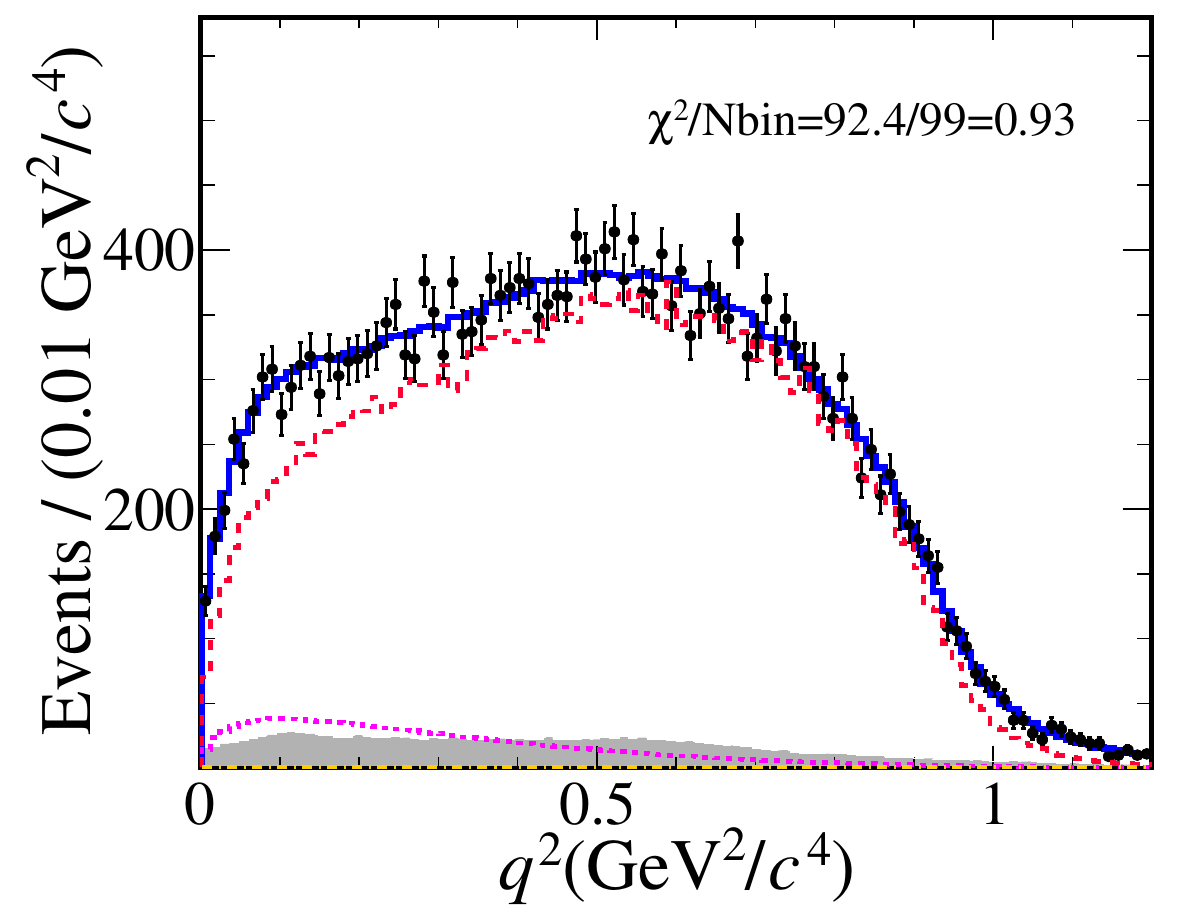} 
    \includegraphics[width=0.23\textwidth]{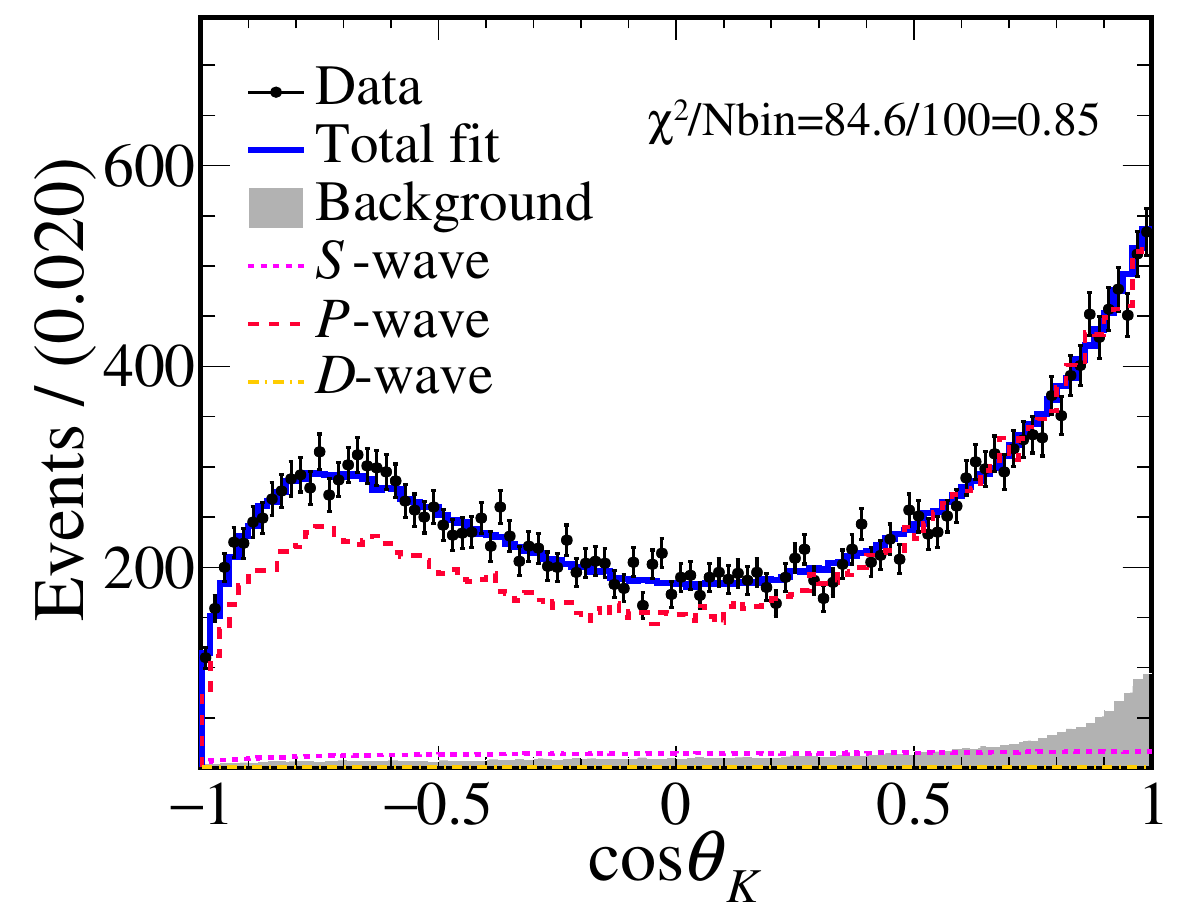}\\[0.1cm]
    \includegraphics[width=0.23\textwidth]{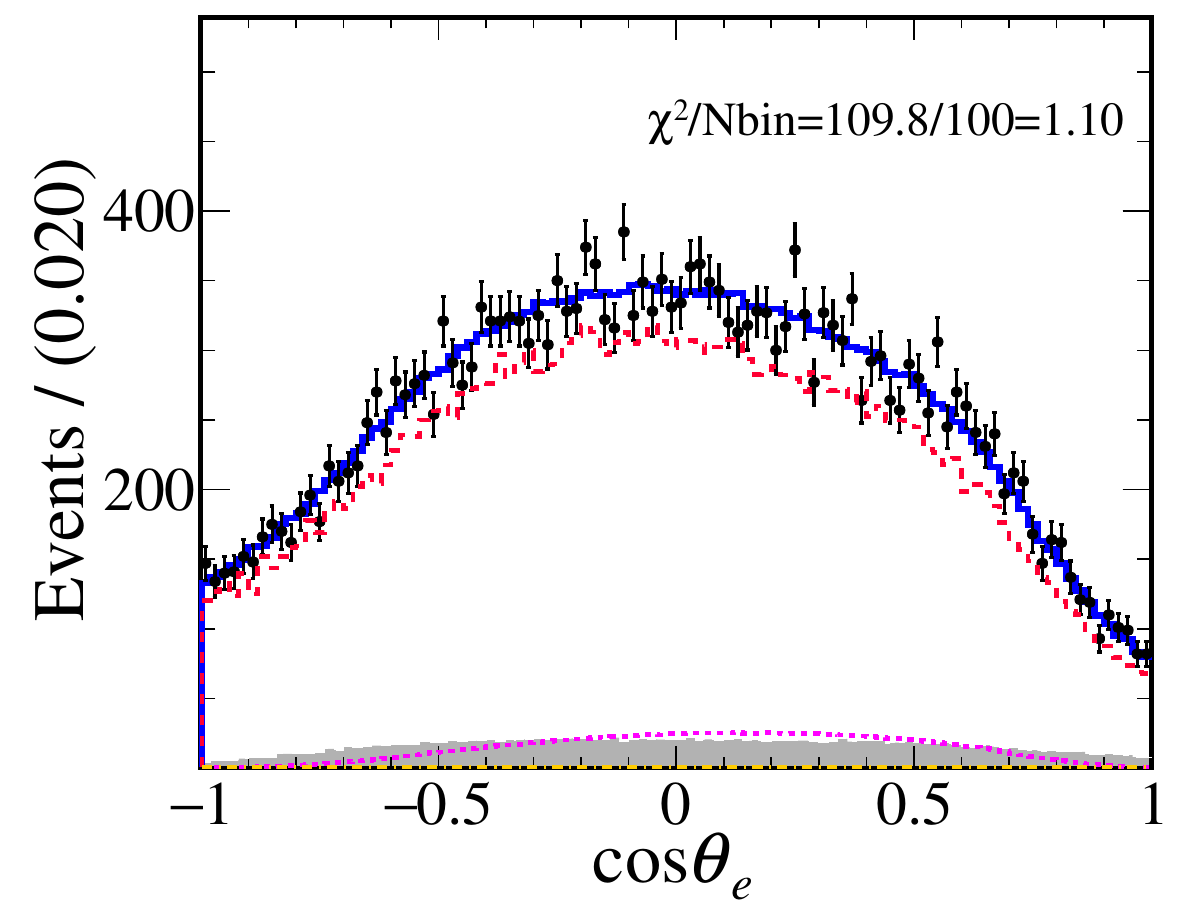} 
    \includegraphics[width=0.23\textwidth]{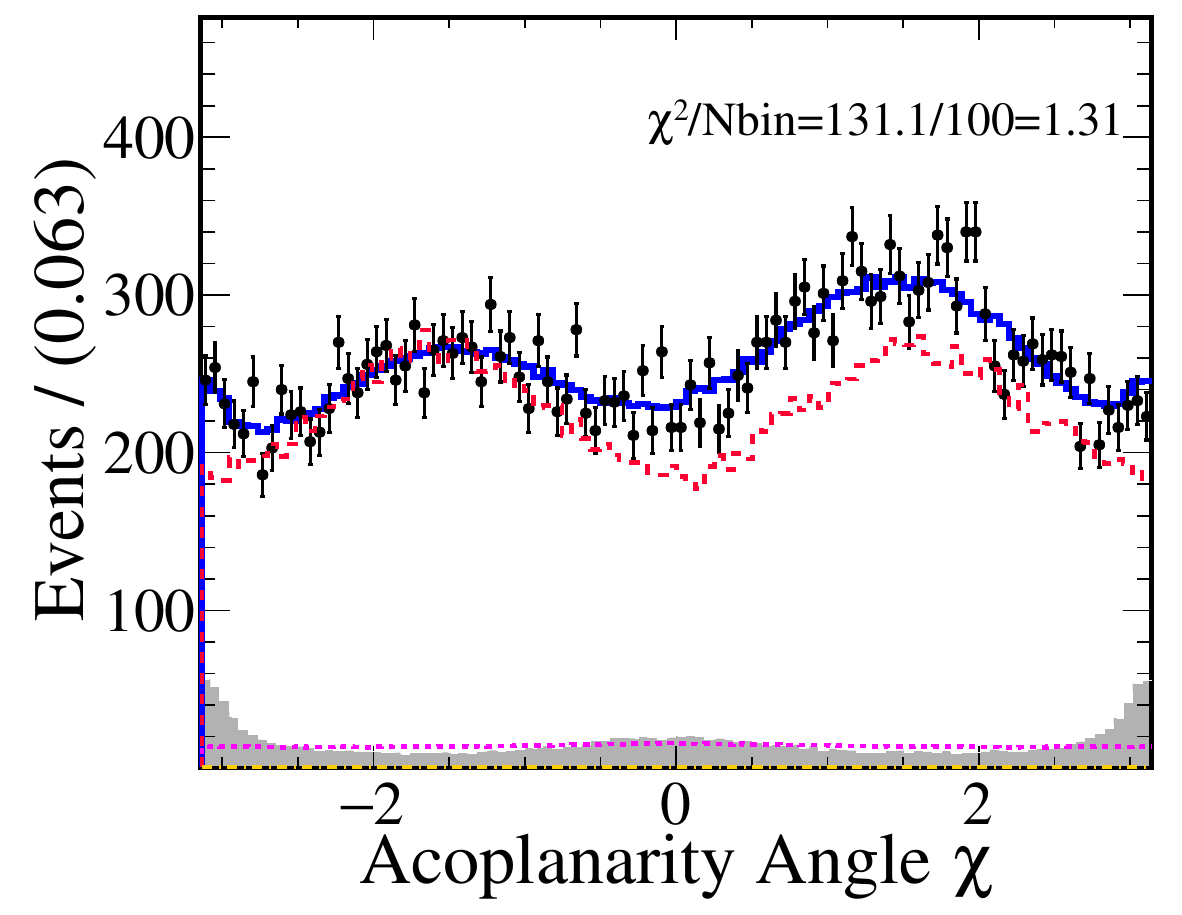}
    \caption{The data and fits for $U_{\rm miss}$ and the projections onto the five kinematic variables of $D^0\to K^-\pi^0 e^+\nu_e $ decay;
    The dots with error bars represent the data, the blue curves indicate the total fit, and the filled gray histograms show the background contributions.    
    The dashed violet, dashed red, and dash-dotted orange lines correspond to the $\mathcal S$, $\mathcal P$, and $\mathcal D$-wave components, respectively. 
    Here ${\rm Nbin}$ denotes the number of bins. For the $\chi^2/$Nbin calculation, we merge neighboring bins with very few entries until at least 20 entries are accumulated. 
    }
    \label{fit:pwa-center}
    \end{figure}

The missing neutrino, $\nu_e$, in the $D^0$ reconstruction can be characterized by the variable $U_{\rm miss} = E_{\rm miss}-|\vec{p}_{\rm miss}|c$, where $E_{\rm miss}$ and $\vec{p}_{\rm miss}$ are the energy and momentum of the missing neutrino inferred from energy and momentum conservation~\cite{BESIIIanffDp2kpienu}, respectively.
To suppress backgrounds, several selection criteria are imposed. The backgrounds of $D^0 \to K^- e^+ \nu_e$ and $D^{0} \to K^- \pi^+ \pi^0$ are rejected by two requirements of $U_{\rm miss}^{\pi^0}>0.04\ \rm GeV$ and $M_{K^-\pi^0 e^+ }<1.70\ \rm GeV/\textit{c}^2$.  Here, $U_{\rm miss}^{\pi^0}$ is analogous to $U_{\rm miss}$ but with the $\pi^0$ candidate excluded from the final state (i.e.,
it peaks at zero for $D^0 \to K^- e^+ \nu_e$ decays) and $M_{K^-\pi^0 e^+ }$ represents the invariant mass of the $K^-\pi^0e^+$ system. A further background  arises due to the exchange of the $\pi^0$ from $D^{-}\to K^+ \pi^- \pi^- \pi^0$ with the $\pi^+$ from $D^{+}\to K^- \pi^+ e^+ \nu_e$.  A quantity $M_{\rm BC}^{\rm new}$, defined for the decay $D^{-}\to K^+ \pi^- \pi^- \pi^0$, is required to lie outside $(1.863,1.877)\ \rm GeV/\textit{c}^{2}$.
After applying all selection criteria, the signal yield is extracted through an unbinned maximum likelihood fit to the $U_{\rm miss}$ distribution. In the fit, the signal is modeled using the signal MC simulated shape, while the background is derived from the inclusive MC sample, and both are convolved with a Gaussian smearing function. All parameters are free in the fit, shown in Fig.~\ref{fit:pwa-center}, and the signal yield is determined to be \Nsig. 

The averaged $D^0$ reconstruction efficiency is \EAsig, including a correction due to the combined discrepancy between data and MC simulation. The BF of $D^0 \to K^- \pi^0 e^+ \nu_e$ is determined to be $(\BFnormal)\times 10^{-3}$ based on the DT method described in Appendix A, where the first uncertainty is statistical and the second is systematic. 
%--------BF sys--------------BF sys---------------BF sys-----------------BF sys
The dominant systematic uncertainties on the BF arise from the $U_{\rm miss}$ fit and efficiency differences between data and MC simulation. By varying the combinatorial background shape, an uncertainty from the $U_{\rm miss}$ fit of 0.3\% is determined.  The uncertainties due to tracking and PID for the $e^+$, 0.14\%, and the $K^-$, 0.14\%, as well as $\pi^0$ reconstruction, 0.2\%, are evaluated using $e^+e^- \to \gamma e^+e^-$ events and DT $D\bar{D}$ hadronic events~\cite{BESIIIDtoKmpi0munu}, respectively. The uncertainty from ST reconstruction, 0.3\%, is from Ref.~\cite{BESIIIDtoKmpi0munu}. The uncertainties associated with the $M_{K^-\pi^0 e^+}$ (0.1\%), $U_{\rm miss}^{\pi^0}$ (0.1\%), and $M_{\rm BC}^{\rm new}$ (0.2\%) requirements are estimated by smearing the signal MC sample.
In the BF calculation, the BF of $\pi^0\to\gamma\gamma$  is taken as $(98.823\pm0.034)\%$~\cite{PhysRevD.110.030001}, with its uncertainty propagated accordingly. 
%The systematic uncertainties are summarized in Ref.~\cite{This-supply}. 
Furthermore,  the uncertainty in MC statistics is determined to be 0.1\%.
The total systematic uncertainty of BF is 0.6\%, from the sum of all contributions in quadrature.

%----------------amplitude analysis-------------------amplitude analysis--------------amplitude analysis-------------
To study the dynamics of $D^0 \to K^- \pi^0 e^+ \nu_e $, the candidate events are further required to be within  $U_{\rm miss} \in(-0.06, 0.06)\ \rm GeV$, while the maximum energy of additional photons $E_{\rm extra\it{\gamma}}^{\rm max}$ unrelated to the $\pi^0$ in both tag and signal sides is required to be less than $0.4\ \rm GeV$. A sample of 25665 candidate events with a purity of $(94.5\pm 0.14)$\% is obtained.

%---------------公式、参数描述-------------公式参数描述--------------公式参数描述---------------
The  differential decay width for $D_{\ell 4}$ decays is given in Ref.~\cite{PhysRevD.83.072001,prd46_5040}, and can be expressed in terms of five kinematic variables $m_{K^-\pi^0}$, $q$, $\cos\theta_{K} $, $\cos \theta_{e}$ and $\chi$\cite{prd46_5040}. The variables $m_{K^-\pi^0}$ and $q$  are the invariant masses of $K^-\pi^0$ and $e^+ \nu_e$, respectively.
There are two helicity angles, the decay angle $\theta_{K}$ between the $K^-$ flight direction in the $K^-\pi^0$ rest frame and the $K^-\pi^0$ system boost direction in the ${D}^0$ frame, and $\theta_{e}$, defined analogously for the $e^+\nu_e$ system. The acoplanarity angle $\chi$ is the orientation between the decay planes of the $K^-\pi^0$ and $e^+ \nu_e$.
The $\mathcal{S}$, $\mathcal{P}$, $\mathcal{P^\prime}$ and \wv{D} contributions are considered in the amplitude analysis.
The \wv{S} phase employs the LASS parameterization~\cite{LASS1988} $\delta_\mathcal{S}(m_{K^-\pi^0})=\delta_{\mathrm{BG}}+\delta_{K^*_0(1430)^-}$, incorporating the contributions from $K_0^*(1430)^-$ and other components. $\delta_{\mathrm{BG}}$ is a function of scattering length $a_{\mathcal{S},\mathrm{BG}}^{1/2}$ and effective range $b_{\mathcal{S},\mathrm{BG}}^{1/2}$ .
The magnitude of the \wv{S}, $\mathcal{A}_\mathcal{S}$, is a function of the dimensionless coefficient $r_\mathcal{S}^{(1)}$ and the relative intensity $r_\mathcal{S}$~\cite{PhysRevD.83.072001}.
The $\mathcal{P}$, $\mathcal{P^\prime}$ and \wv{D} amplitudes are described by the Breit-Wigner functions with a parameter $r_{\mathrm{BW}}$, which is the effective radius of the strong-interaction barrier, and the masses and widths of the $K\pi$ resonances~\cite{PhysRevD.83.072001}. 
Moreover, the helicity FFs can in turn be related to the two axial-vector FFs $A_1(q^2)$, $A_2(q^2)$, and a vector FF $V(q^2)$ for the $\mathcal{P, P^\prime}$ waves. The FFs of the \wv{D} also include two axial-vector and one vector FFs, $T_1(q^2)$, $T_2(q^2)$ and $T_V(q^2)$.  All of them are parametrized with the simple pole form, with pole masses $m_A$ and $m_{V}$. At $q^2=0$, the FF ratios $r_V=V(0)/A_1(0)$ and $r_2=A_2(0)/A_1(0)$ are determined from the fit.
For the $K^*_2(1430)^-$, only the amplitude magnitude $r_{\mathcal{D}}$ and phase $\phi_{\mathcal{D}}$ are floated, while the FFs $r_{2V}=T_V(0)/T_1(0)$ and $r_{22}=T_2(0)/T_1(0)$ are fixed to 1~\cite{Leibovich:1997em, Charles:1998dr}, and the masses and widths of $K^*(1410)^-$ and $K^*_2(1430)^-$ are fixed to their PDG values~\cite{PhysRevD.110.030001}.

%---------本底参数化--------本底参数化--------本底参数化----------------------------
The amplitude analysis is performed on the differential decay width using the unbinned maximum likelihood fit method.  The negative log-likelihood function is defined as
\begin{align}
  \label{eq:loglike}
  -\ln \mathcal{L} &= -\sum_{i=1}^{N}\left[(1-f_{b}) \, \frac{\omega(\xi,\eta)}{\int d\xi\, \omega(\xi,\eta) \epsilon(\xi)R_4(\xi_i)}\right.\nonumber\\
  &\left.+\,f_{b} \, \frac{B_{\epsilon}(\xi)}{\int d\xi\, B_{\epsilon}(\xi)\epsilon(\xi)R_4(\xi_i)}\right],
\end{align}
where $f_b$ is the background fraction obtained from the $U_{\rm miss}$ fit, $\omega(\xi,\eta)$ represents the signal probability density function (PDF) described by the intensity distribution~\cite{BESIIIanffDp2kpienu}, and $B_\epsilon(\xi)$ denotes the efficiency-corrected background PDF described by the inclusive MC sample. The efficiency $\epsilon(\xi)$ is evaluated using the signal MC sample generated in phase space, by binning the five-dimensional variables and calculating the efficiency in each bin, and $R_4(\xi_i)$ is an element of four-body phase space. Here, $\xi$ and $\eta$ correspond to the five-dimensional kinematic variables and the fit parameters, respectively. The integrals in the denominators are performed using the same signal MC samples used for signal and background PDFs.
%serve as normalization constants, . 
%Phase-space MC is used for signal significance calculations, while signal MC is employed for nominal solutions.

%-------显著度----------显著度----------------显著度----------------
The statistical significances of the $\mathcal{D}$  and $\mathcal{P^\prime}$ waves are evaluated via the likelihood ratio $\Lambda_\mathcal{D}=-2\ln{(\mathcal{L}_\mathcal{SP}/\mathcal{L}_\mathcal{SPD}})$ and $\Lambda_{P^\prime}=-2\ln{(\mathcal{L}_\mathcal{SPD}/\mathcal{L}_\mathcal{SPDP^\prime}})$, respectively,  where $\mathcal{L}_{\mathcal{SP}}$, $\mathcal{L}_{\mathcal{SPD}}$, and $\mathcal{L}_{\mathcal{SPDP^\prime}}$ denote the maximum likelihood values obtained by including the \wv{S} and \wv{P} components only, with the \wv{D} added, and with both \wv{D} and $\mathcal{P^\prime}$ waves added, respectively.
Assuming that $\Lambda_{\mathcal{D}}$ follows a $\chi^2$ distribution with two degrees of freedom and one degree of freedom for $\Lambda_{\mathcal{P^\prime}}$, the ${\mathcal D}$ and \wv{P^\prime} components have significances of $7.9\sigma$ and $2.5\sigma$, respectively.
Consequently, the amplitude analysis including only the $\mathcal{S}$, $\mathcal{P}$, and $\mathcal{D}$-wave components is taken as the nominal solution.

%--------------中心解----------中心解------------中心解------------
The nominal fit results are summarized in Table~\ref{tab:ff-sum}, with projections of the five-dimensional variables shown in Fig.~\ref{fit:pwa-center}. The fractions of the $\mathcal{S}$, $\mathcal{P}$ and \wv{D} components ($f_{\mathcal{S,P,D}}$) are shown in Table~\ref{tab:ff-sum}.  The amplitudes for waves of different spin are orthogonal and do not interfere in the total rate. The corresponding BFs are \textcolor{black}{${\mathcal B}$($D^0\to (K^-\pi^0)_{\mathcal{S}\text{-}{\rm wave}}~e^+\nu_e$)  = $(\BFSwave)\times 10^{-3}$},  \textcolor{black}{${\mathcal B}$($D^0\to(K^-\pi^0)_{\mathcal{P}\text{-} {\rm wave}}~ e^+\nu_e =(\BFPwave)\times 10^{-3}$} and \textcolor{black}{${\mathcal B}$($D^0\to (K^-\pi^0)_{\mathcal{D}\text{-} {\rm wave}}~ e^+\nu_e) =(\BFDwave)\times 10^{-5}$}. The correlation matrix for the fitted parameters is provided in Ref.~\cite{This-supply}.

\begin{table}[htp]
  \begin{center}
  \caption{ \textcolor{black}{ The parameters of the nominal solution, where the first and second uncertainties are statistical and systematic, respectively.} No systematic uncertainty is reported for $m_{K^*(892)^-}$ due to the lack of a suitable energy calibration. }
  \vspace{0.25cm}
  \begin{tabular}{l|c}
  \hline
  \hline
  Variable                                 & Value  \\  
  \hline
$r_\mathcal{S}$(GeV)$^{-1}$& $-7.53\pm0.22\pm0.11$\\
$r_\mathcal{S}^{(1)}$& $~~~~0.14\pm0.04\pm0.01$\\
$a^{1/2}_{\mathcal{S},\mathrm{BG}}~(\rm GeV/\textit{c})^{-1}$& $~~~~1.98\pm0.10\pm0.13$\\
$b^{1/2}_{\mathcal{S},\mathrm{BG}}~(\rm GeV/\textit{c})^{-1}$ & $~~~~0.57\pm0.53\pm0.27$ \\
$m_{A}(\rm GeV/\it{c}^{2}) $& $~~~~2.72\pm0.18\pm0.11$ \\
$m_{V}(\rm GeV/\it{c}^{2}) $& $~~~~1.70\pm0.11\pm0.02$ \\
$r_{V}$ & $~~~1.41\pm0.05\pm0.01$ \\
$r_2$ & $~~~0.77\pm0.04\pm0.02$ \\
$m_{K^{*}(892)^{-}}(\rm MeV/\it{c}^{2})$& $892.9\pm0.2$\\
$\Gamma_{K^{*}(892)^{-}}^{0}(\rm MeV)$& $47.9\pm0.5\pm0.4$\\
$r_\mathrm{BW}~(\rm GeV/\textit{c})^{-1}$ & $~~~~3.38\pm0.17\pm0.16$ \\
$r_{\mathcal{D}}$(GeV)$^{-4}$ & $~~~11.0\pm1.6\pm1.7$ \\
$\phi_{\mathcal{D}}$ (degree)& $-16.9\pm7.7\pm3.0$ \\
$f_{\mathcal{S}}$  (\%)& $~~~5.86\pm0.18\pm0.21$\\
$f_{\mathcal{P}}$ (\%)& $~93.97\pm0.19\pm0.21$ \\
$f_{\mathcal{D}}$  (\%)& $~~~0.16\pm0.05\pm0.02$ \\
  
  \hline
  \hline
  \end{tabular}
  \label{tab:ff-sum}
  \end{center}
  \end{table}
%%%%%%%%%%%%%%%%%%%%%%%%%%%%%%%%%%
%------------A1(0)----------------A1(0)---------------------A1(0)--------------------A1(0)-------------
We also obtain the product of $|V_{cs}| A_1(0)=0.618\pm 0.010_{\rm stat}\pm 0.008_{\rm syst }$  based on the nominal solution. The details can be found in Appendix B. Taking the $|V_{cs}|=0.97345 \pm 0.00016$ from the global SM fit \cite{PhysRevD.110.030001}, we find $A_1(0) = 0.634 \pm 0.010_{\rm stat} \pm 0.009_{\rm syst}$.

\begin{figure}[htp]
\centering
\includegraphics[width=0.23\textwidth]{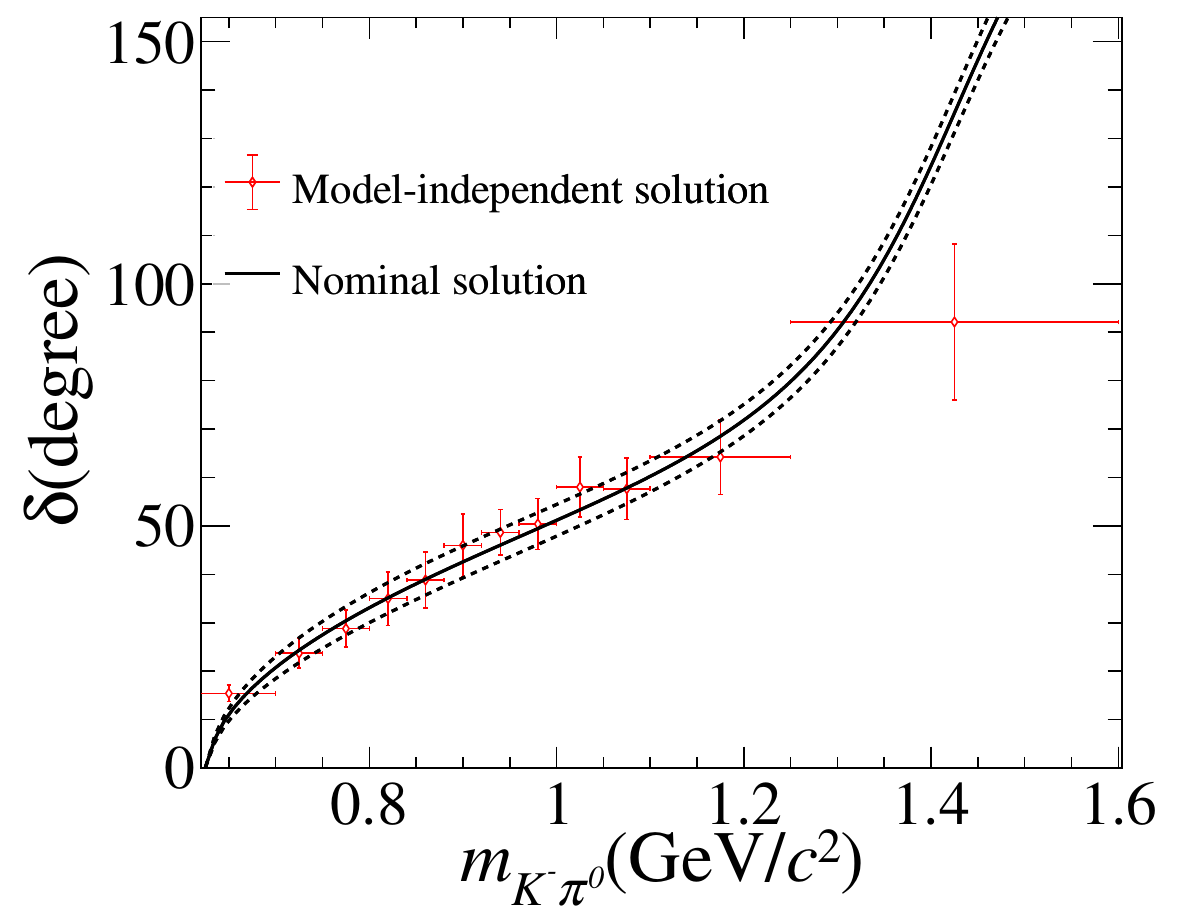}
\includegraphics[width=0.23\textwidth]{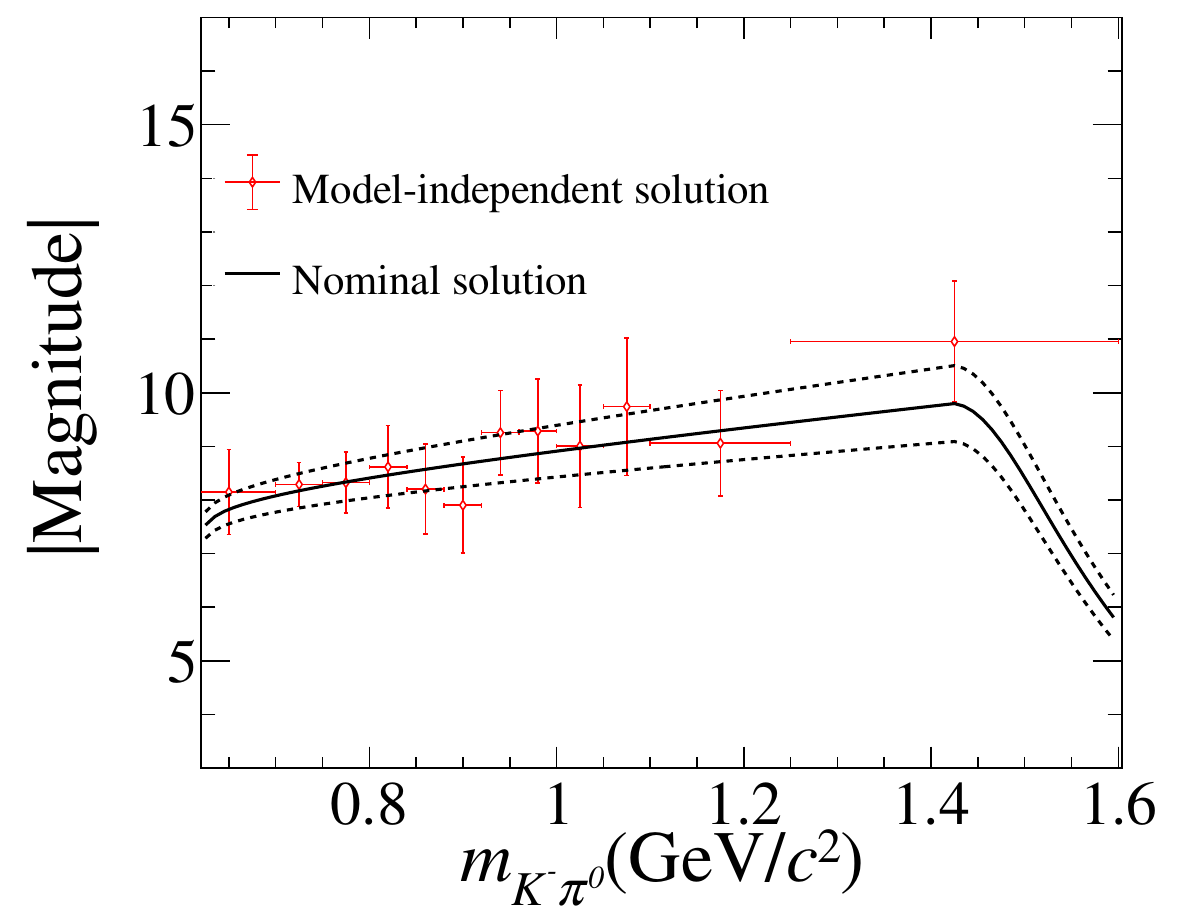}
\caption{\textcolor{black}{\wv{S} phase (left) and magnitude (right) versus $m_{K ^-\pi^0}$, assuming that the signal is composed of the $\mathcal{S}$, $\mathcal{P}$ $\left({K}^{*}(892)^-\right)$ and \wv{D} $\left({K}_2^*(1430)^-\right)$. The dots with error bars represent the fit results from the model-independent measurement, and the black solid line corresponds to the nominal solution based on Table~\ref{tab:ff-sum}, while the dashed lines reflect the propagated $\pm 1\sigma$ uncertainties.
}}
\label{fig:deltaS_m}
\end{figure}

The nominal parametrization of the \wv{S} includes only the $K^*_0(1430)^-$ resonance explicitly. However, the lighter scalar resonance $K^*_0(700)$ is also expected to contribute to the $K\pi$ system \cite{Zhou-2006wm,Descotes-Genon-2006sdr,kappaLQCD2012}. Since the $K^*_0(700)$ remains a long-standing puzzle, a model-independent determination of the \wv{S} amplitude is particularly valuable.
To get the phase shift and magnitude of the \wv{S} in a model-independent way and validate its nominal parameterization model, we perform another fit by dividing the $m_{K^-\pi^0}$ spectrum into 12 bins with free phases and magnitudes of the \wv{S}. The results are illustrated in Fig.~\ref{fig:deltaS_m}; they are consistent with the nominal solution. Details are given in Ref.~\cite{This-supply}. 

%--------------PWA sys-------------PWA sys--------------------PWA sys-----------------
The systematic uncertainties in the amplitude analysis are estimated by comparing the nominal solution to those obtained by varying relevant input parameters or conditions within their uncertainties.
First, the fixed parameters $r_{2V}$, $r_{22}$, as well as the mass and width of the $K^*_2(1430)^-$ are varied by $\pm0.5$ and  $\pm1\sigma$~\cite{PhysRevD.110.030001}, respectively. The background fraction $f_b$ and the background shapes including $e^+e^- \to q\bar{q}$ and $e^+e^- \to D\bar{D}$ are also varied by $\pm 1\sigma$ based on the $U_{\rm miss}$ fit and Refs.~\cite{BESIIICrossDD,BESIIICrossNonDD,BESIIICrossQQ}, respectively.
Corrections related to PID and tracking efficiencies are applied to the signal MC sample, and the resulting changes in the fits are taken as the systematic uncertainties. The relative values can be found in Ref.~\cite{This-supply}.

\begin{figure}[htp]
  \centering
  \includegraphics[width=0.4\textwidth]{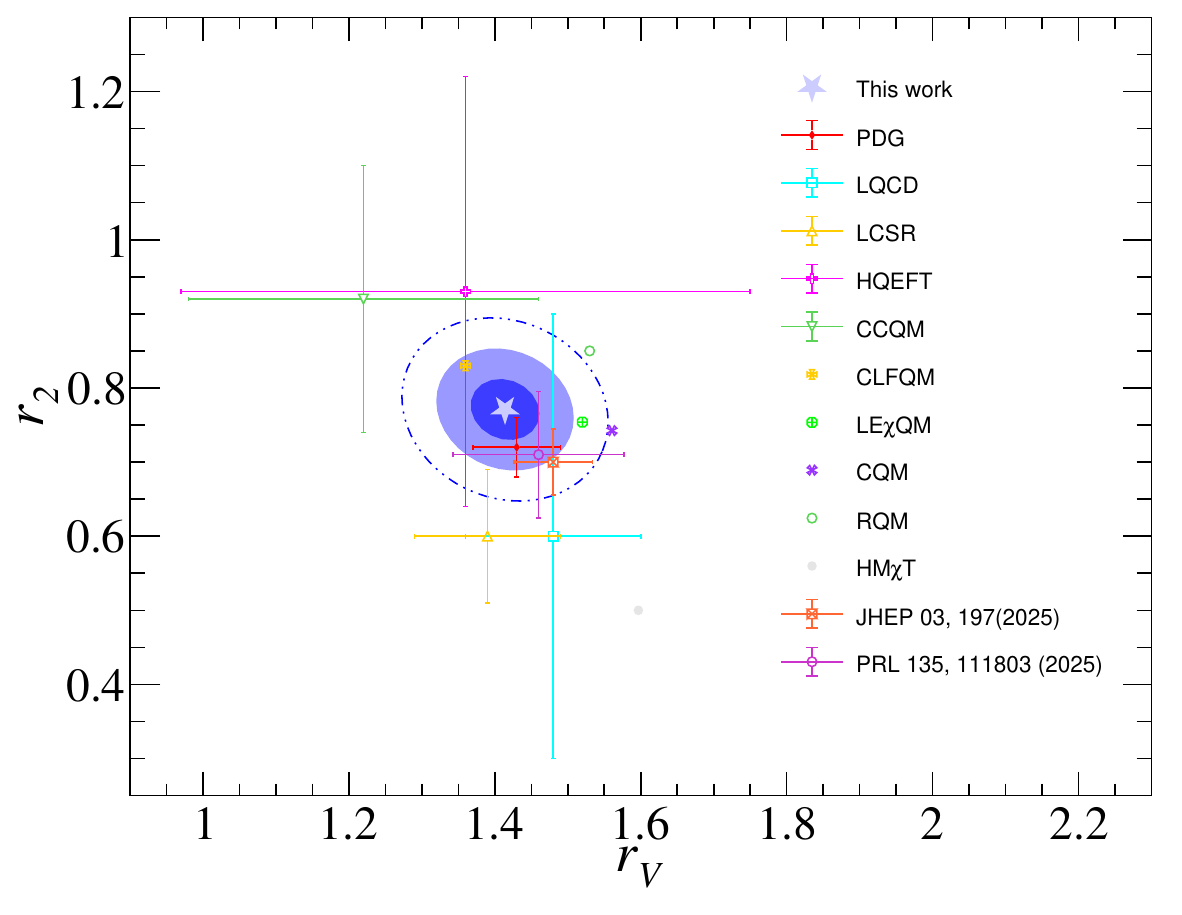}
  \caption{Comparison of the measured FFs, PDG values and theoretical predictions from Refs.~\cite{HMChiT2005,CLFQM2012,ABADA2003625,CCQM2019,CQM2017,CQM2000,LCSR2006,LCSR2020,RQM2020,CLFQM2012,CLFQM2017,LEChiQM2014,RQM2020,HQCD2024,HQEFT2003}. The inner, middle, and outer ellipses correspond to uncertainty ranges of $1\sigma$, $2\sigma$ and $3\sigma$ deviations, respectively.
  }
  \label{fig:CFF}
  \end{figure}

%----------Summary-------------Summary-----------------Summary---------------------
In summary, this Letter presents the first amplitude analysis of $D^0\to K^-\pi^0e^+\nu_e$ by analyzing $e^+e^-$ annihilation data corresponding to an integrated luminosity of 20.3 fb$^{-1}$ collected at $\sqrt{s}=$ 3.773 GeV with the BESIII detector. 
The $K^*_2(1430)^-$ component is observed for the first time with a significance of $7.9\sigma$. The BF is measured to be $\mathcal{B}(D^0 \to K_2^*(1430)^- e^+ \nu_e) = (\BFKstarD)\times10^{-5}$, where $\mathcal{B}(K_2^*(1430)^- \to (K\pi)^-) = (49.9 \pm 1.2)\%$ is taken from Ref.~\cite{PhysRevD.110.030001}, assuming isospin symmetry. This result is consistent within $3\sigma$ with the predictions based on SU(3) flavor symmetry~\cite{BFKstar2WRM} and the RQM~\cite{TensorRQM}.
In addition, the FFs $r_V$ and $r_2$ of the $D^0 \to K^*(892)^- $ transition are measured with high precision. A comparison with the PDG average ~\cite{PhysRevD.110.030001}, recent BESIII measurements~\cite{BESIIID02kspimunu,BESIIILiLeiDtoKpienu} and various theoretical calculations is shown in  Fig.~\ref{fig:CFF}. These results help to discriminate between models, but the lack of imprecise meaning of theoretical uncertainties make precise conclusions difficult.  
The product of $|V_{cs}|A_1(0)$ is measured to be $0.619\pm 0.010_{\rm stat}\pm 0.008_{\rm syst }$,
which can provide an independent determination of the $|V_{cs}|$ to improve its precision with the theoretical input $A_1(0)$~\cite{CKMDToV2025}. 
Combining the BF of \textcolor{black}{${\cal B}(D^0 \rightarrow K^{*}(892)^-  \mu^+ \nu_{\mu},K^{*}(892)^-\to K^- \pi^0) =(6.87 \pm 0.13_{\rm stat} \pm 0.11_{\rm syst})\times 10^{-3}$} from the previous BESIII measurement~\cite{BESIIIDtoKmpi0munu}, we obtain the ratio $\mathcal{R}_{\rm LFU} = \mathcal{B}(D^0\to K^*(892)^-\mu^+ \nu_\mu)/\mathcal{B}(D^0\to K^*(892)^-e^+\nu_e) = 0.928\pm0.020_{\rm stat}\pm0.012_{\rm syst}$, where correlated systematic uncertainties have been canceled. 
This result is compatible at 2.7\% precision level with the theoretical predictions in the SM 0.921-0.945~\cite{LCSR2006,LCSR2025,CQM2017,CCQM2019}.
Furthermore, by combining with \textcolor{black}{${\cal B}(D^0 \rightarrow K^{*}(892)^-  e^+ \nu_{e},K^{*}(892)^-\to K^0_S \pi^-) =(6.80 \pm 0.10_{\rm stat} \pm 0.11_{\rm syst})\times 10^{-3}$} from the previous BESIII measurement~\cite{BESIIILiLeiDtoKpienu}, we obtain $\mathcal R_{K^{*-}} =\mathcal{B}(K^*(892)^-\to K^- \pi^0)/\mathcal{B}(K^*(892)^-\to K_S^0 \pi^-) = 1.09\pm0.02_{\rm stat}\pm0.02_{\rm syst}$. 
This ratio is consistent with the value $\mathcal R_{K^{*-}} = 1.07 \pm 0.02_{\rm stat} \pm 0.03_{\rm syst}$, derived from the measurements in the muon channel~\cite{BESIIIDtoKmpi0munu,BESIIID02kspimunu}. These ratio values indicate that the isospin-breaking effect in the $K^*(892)^-$ decay may need to be reconsidered, assuming the maximal breaking effect with a few percent. We also extract the phase shift of the \wv{S} with the $m_{K^-\pi^0}$ in a model-independent way, which provides unique insight into the nature of $K^*_0(700)$.

%\end{CJK}

%% Saved at => 2025-07-21
\textbf{Acknowledgement}

The BESIII Collaboration thanks the staff of BEPCII (https://cstr.cn/31109.02.BEPC) and the IHEP computing center for their strong support. This work is supported in part by National Key R\&D Program of China under Contracts Nos. 2023YFA1606000, 2023YFA1606704; National Natural Science Foundation of China (NSFC) under Contracts Nos. 12475092, 12165022, 11635010, 11935015, 11935016, 11935018, 12025502, 12035009, 12035013, 12061131003, 12192260, 12192261, 12192262, 12192263, 12192264, 12192265, 12221005, 12225509, 12235017, 12361141819; Yunnan Fundamental Research Project under Contract No. 202301AT070162; the Chinese Academy of Sciences (CAS) Large-Scale Scientific Facility Program; the Strategic Priority Research Program of Chinese Academy of Sciences under Contract No. XDA0480600; CAS under Contract No. YSBR-101; 100 Talents Program of CAS; The Institute of Nuclear and Particle Physics (INPAC) and Shanghai Key Laboratory for Particle Physics and Cosmology; ERC under Contract No. 758462; German Research Foundation DFG under Contract No. FOR5327; Istituto Nazionale di Fisica Nucleare, Italy; Knut and Alice Wallenberg Foundation under Contracts Nos. 2021.0174, 2021.0299; Ministry of Development of Turkey under Contract No. DPT2006K-120470; National Research Foundation of Korea under Contract No. NRF-2022R1A2C1092335; National Science and Technology fund of Mongolia; Polish National Science Centre under Contract No. 2024/53/B/ST2/00975; STFC (United Kingdom); Swedish Research Council under Contract No. 2019.04595; U. S. Department of Energy under Contract No. DE-FG02-05ER41374

%% ends here %%
%\bibliographystyle{plain}  
\bibliography{D02kpienu_draft} 

\appendix
%\onecolumngrid
%\section{Appendix}

\subsection{Appendix~A: Branching fraction in DT method}
\label{sec:appA}

Based on the DT method, the BF of the $D^0\rightarrow K^{-}\pi^0e^+\nu_e$ decay is expressed as
\begin{equation}
\mathcal{B}_{\rm SL}=\frac{N_{\rm DT}}{N^{\rm tot}_{\rm ST} \, \bar\epsilon_{\rm SL}}, \label{eq:branch}
\end{equation}where $N_{\rm DT}$ and $N^{\rm tot}_{\rm ST}=\sum^{3}_{i=1} N^{i}_{\rm ST}$   are the total DT and ST yields summing over all tag modes, respectively.  $\bar\epsilon_{\rm SL}=\sum^{3}_{i=1} \frac{N^{i}_{\rm ST}}{N^{\rm tot}_{\rm ST}} \,\frac{\epsilon^i_{\rm DT}}{\epsilon^i_{\rm ST}}$
is the averaged efficiency of reconstructing the $D^0\rightarrow K^{-}\pi^0e^+\nu_e$ decay,  where  $\epsilon^i_{\rm DT}$ and $\epsilon^i_{\rm ST}$
are the DT and ST efficiencies for the $i^{\rm th}$ tag mode, respectively.

\subsection{Appendix~B: Extracting the product of $|V_{cs}|A_1(0)$}
\label{sec:appB}

The differential decay width of the \wv{P} $K^*(892)^-$ is integrated over three angles yielding 
\begin{equation}
\begin{aligned}
  \frac{d \Gamma}{d q^2 d m^2_{K^-\pi^0}} = &\frac{1}{3} \frac{G_F^2\left|V_{c s}\right|^2}{(4 \pi)^5 m^2_{D^0}} \beta p_{K^- \pi^0}  \\ 
  & \left[\frac{2}{3}\left\{\left|\mathcal{F}_{11}\right|^2+\left|\mathcal{F}_{21}\right|^2+\left|\mathcal{F}_{31}\right|^2\right\}\right].
\label{eq:Gamma_q2m2}
\end{aligned}
\end{equation}
In this expression, $G_F$ is the Fermi coupling constant, $m_D$ is the mass of $D^0$ meson, $V_{cs}$ is the $c\to s$ CKM matrix element,  $\mathcal F_{i1}$ are the FF terms corresponding to the $\mathcal P$ wave component, and $\beta=2p^*/m_{K^-\pi^0}$, where $p^*$ is the momentum of $K^-$ in the $K^-\pi^0$ rest frame and  $p_{K^-\pi^0} $ is the momentum of the $K^-\pi^0 $system in the $D^0 $ rest frame. 

Replacing The $\mathcal F_{i1}$ in Eq.(\ref{eq:Gamma_q2m2}) with helicity FFs transforms this decay width into 
\begin{equation}
\begin{aligned}
  \frac{d \Gamma}{d q^2 d m^2_{K^-\pi^0}} = & \frac{G_F^2\left|V_{c s}\right|^2}{96 \pi^4 m^2_{D^0}} \frac{p^* \mathcal{B}_{K^*}}{m_{K^-\pi^0}} \frac{p_{K \pi} q^2}{p_0^* \Gamma_0} \\
  & \left(\left|H_0\right|^2+\left|H_{+}\right|^2+\left|H_{-}\right|^2\right)|\mathcal{A}(m_{K^-\pi^0})|^2.
\label{eq:Gamma_q2m2_x}
\end{aligned}    
\end{equation}
Here, $\mathcal{B}_{K^*}=0.354\pm0.006$ denotes the BF of $K^*(892)^-\to K^-\pi^0$ considering the ratio $\mathcal{R}_{K^*}$ measured in this Letter.
The amplitude-squared term $|\mathcal{A}(m_{K^-\pi^0})|^2$ describing the $K^*(892)^- $ is given by
\begin{equation}
|\mathcal{A}(m_{K^-\pi^0})|^2=\frac{\left(m_0 \Gamma_0 F_1(m_{K^-\pi^0})\right)^2}{\left(m_0^2-m^2_{K^-\pi^0}\right)^2+\left(m_0 \Gamma(m_{K^-\pi^0})\right)^2}
\label{eq:Amplitude2},
\end{equation}
where $m_0$ and $\Gamma_0 $ are the mass and width of $K^*(892)^- $, respectively, $F_1(m_{K^-\pi^0})=(\frac{p^*}{p_0^*}) \frac{B_1(p^*)}{B_1(p_0^*)}$ with the Blatt-Weisskopf damping factor $B_1(p^*)=1/\sqrt{1-r_\mathrm{BW}p^*}$.
The decay width is 
\begin{equation}
{\Gamma} = \frac{\hbar \mathcal{B}\left(D^{0} \rightarrow K^*(892)^- e^{+} \nu_e\right) \mathcal{B}\left(K^*(892)^- \rightarrow K^{-} \pi^{0}\right)}{\tau_{D^{0}}},
\end{equation}
where $\tau_{D^0}$ is the lifetime of $D^0$ meson, $\hbar$ is the Planck constant, and the BF product is measured in this analysis.

Finally, by constraining to the decay width, a two-dimensional integration of Eq.~(\ref{eq:Gamma_q2m2_x}) is performed with the parameters from the nominal solution to extract the product of $|V_{cs}|A_1(0)$. The systematical are mainly produced from the amplitude analysis solution and the uncertainty from PDG value.

\onecolumngrid
\section*{}
\begingroup
\small
\renewcommand{\author}[1]{\begin{center}#1\end{center}}
%% Saved at => 2025-07-21
\author{
M.~Ablikim$^{1}$\BESIIIorcid{0000-0002-3935-619X},
M.~N.~Achasov$^{4,b}$\BESIIIorcid{0000-0002-9400-8622},
P.~Adlarson$^{81}$\BESIIIorcid{0000-0001-6280-3851},
X.~C.~Ai$^{86}$\BESIIIorcid{0000-0003-3856-2415},
R.~Aliberti$^{38}$\BESIIIorcid{0000-0003-3500-4012},
A.~Amoroso$^{80A,80C}$\BESIIIorcid{0000-0002-3095-8610},
Q.~An$^{77,63,\dagger}$,
Y.~Bai$^{61}$\BESIIIorcid{0000-0001-6593-5665},
O.~Bakina$^{39}$\BESIIIorcid{0009-0005-0719-7461},
Y.~Ban$^{49,g}$\BESIIIorcid{0000-0002-1912-0374}
% ...
\\
\vspace{0.2cm}
(BESIII Collaboration)\\
\vspace{0.2cm} {\it
$^{1}$ Institute of High Energy Physics, Beijing 100049, People's Republic of China\\
% ...
$^{4}$ Budker Institute of Nuclear Physics SB RAS (BINP), Novosibirsk 630090, Russia\\
% ...
$^{38}$ Johannes Gutenberg University of Mainz, Johann-Joachim-Becher-Weg 45, D-55099 Mainz, Germany\\
$^{39}$ Joint Institute for Nuclear Research, 141980 Dubna, Moscow region, Russia\\
% ...
$^{49}$ Peking University, Beijing 100871, People's Republic of China\\
% ...
$^{61}$ Southeast University, Nanjing 211100, People's Republic of China\\
% ...
$^{63}$ State Key Laboratory of Particle Detection and Electronics, Beijing 100049, Hefei 230026, People's Republic of China\\
% ...
$^{77}$ University of Science and Technology of China, Hefei 230026, People's Republic of China\\
% ...
$^{80}$ University of Turin and INFN, (A)University of Turin, I-10125, Turin, Italy; (B)University of Eastern Piedmont, I-15121, Alessandria, Italy; (C)INFN, I-10125, Turin, Italy\\
$^{81}$ Uppsala University, Box 516, SE-75120 Uppsala, Sweden\\
% ...
$^{86}$ Zhengzhou University, Zhengzhou 450001, People's Republic of China\\

\vspace{0.2cm}
$^{\dagger}$ Deceased\\
% ...
$^{b}$ Also at the Novosibirsk State University, Novosibirsk, 630090, Russia\\
% ...
$^{g}$ Also at State Key Laboratory of Nuclear Physics and Technology, Peking University, Beijing 100871, People's Republic of China\\
% ...

}
}

%% ends here %%

\endgroup
\twocolumngrid
\maketitle
\end{document}